\documentclass [conference, letterpaper]{IEEEtran}
\IEEEoverridecommandlockouts
\usepackage{cite}
\usepackage{amsmath,amssymb,amsfonts}
\usepackage{algorithm}
\usepackage{algpseudocode}
\usepackage{graphicx}
\usepackage{tabularx}
\usepackage{textcomp}
\usepackage{xcolor}
\usepackage{subfigure}
\usepackage[letterpaper, top=1.91cm, bottom=2.7cm,
left=1.92cm, right=1.92cm] {geometry}
\begin{document}

\title{Routing and Wavelength Assignment with Minimal Attack Radius for
QKD Networks}
\author{
\IEEEauthorblockN{Mengyao Li\IEEEauthorrefmark{1}, Qiaolun Zhang\IEEEauthorrefmark{1}, Zongshuai Yang\IEEEauthorrefmark{1}, Stefano Bregni\IEEEauthorrefmark{1}, Alberto Gatto\IEEEauthorrefmark{1}, \\ 
Raouf Boutaba\IEEEauthorrefmark{2}, 
Massimo Tornatore\IEEEauthorrefmark{1}}
\IEEEauthorblockA{
    \IEEEauthorrefmark{1}\textit{Politecnico di Milano, Italy} \quad 
     \IEEEauthorrefmark{2}\textit{University of Waterloo, Canada} \\
    %
    Corresponding author: qiaolun.zhang@polimi.it
}
}
\maketitle
\begin{abstract}
Quantum Key Distribution (QKD) can distribute keys with guaranteed security but remains susceptible to \textbf{key exchange interruption} due to physical-layer threats, such as high-power jamming attacks. To address this challenge, we first introduce a novel metric, namely Maximum Number of Affected Requests (maxNAR), to quantify the worst-case impact of a single physical-layer attack, and then we investigate a new problem of Routing and Wavelength Assignment with Minimal Attack Radius (RWA-MAR). 
We formulate the problem using an Integer Linear Programming (ILP) model and propose a scalable heuristic to efficiently minimize maxNAR. Our approach incorporates key caching through Quantum Key Pools (QKPs) to enhance resilience and optimize resource utilization. Moreover, we model the impact of different QKD network architectures, employing Optical Bypass (OB) for optical switching of quantum channels and Trusted Relay (TR) for secure key forwarding. 
 Moreover, a tunable parameter is designed in the heuristic to guide the preference for OB or TR, offering enhanced adaptability and dynamic control in diverse network scenarios. Simulation results confirm that our method significantly outperforms the baseline in terms of security and scalability.

\end{abstract}


\vspace{-3mm}
\section{Introduction}
Quantum Key Distribution (QKD) allows to exchange cryptographic keys with guaranteed information-theoretic security, relying on the laws of quantum mechanics to ensure resilience against quantum-capable adversaries \cite{cao2022evolution, chen2021integrated}. Initially applied in point-to-point configurations, recent advances have enabled the development of more scalable multi-point QKD networks over optical infrastructures \cite{cao2018time}. 
These scalable QKD networks, composed of interconnected QKD nodes, links, and QKD modules, are now transitioning from experimental systems to practical deployments, which can protect highly sensitive domains such as financial transactions and defense communications \cite{sena2024deploying}.

To support these mission-critical applications, prior research has focused on optimizing network performance, particularly through efficient resource provisioning and routing strategies \cite{cao2022evolution, zhang2023routing}. However, comparatively little attention has been paid to the resilience of QKD networks against targeted attacks. Various threat vectors can disrupt key distribution, including out-of-band attacks that interfere with classical control channels or inject unauthorized signals to compromise quantum channels \cite{furdek2011physical,smith2021out, alomari2024securing}. Such attacks are the key reasons why important institutions (e.g. NSA, NIST) remain skeptical about the practicality of QKD \cite{renner2023debate, beullens2022breaking}. 
However, our approach offers a concrete step toward mitigating these vulnerabilities.
For instance, an adversary may utilize high-power jamming malicious signals to attack a single quantum link, which may disrupt not only the intended transmission but also all other requests sharing the same fiber. The impact of the above-mentioned attacks is closely related to the technologies used for QKD networking as discussed below.  

\begin{figure}[htbp]
\centering
\includegraphics[width=0.45\textwidth]{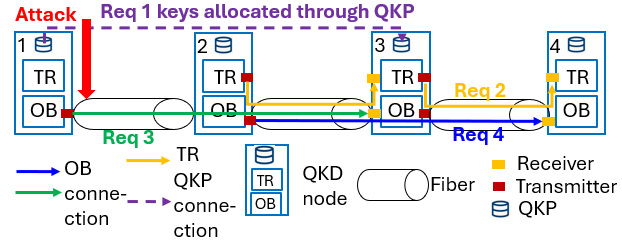}
\setlength{\abovecaptionskip}{-1mm}
\caption{Example of QKD-specific technologies}
\label{fig:keyallocation}
\end{figure}

In this work, we consider QKD networks equipped with three critical technologies: (i) Quantum Key Pools (QKPs), which act as key caches at each node, storing
pre-distributed keys for future use; 
(ii) Trusted Relays (TR), which can forward keys, in case maximum signal reach is insufficient, through intermediate, trusted nodes using one-time pad encryption schemes \cite{cao2022evolution}; and (iii) Optical Bypass (OB), enabling direct key delivery between non-adjacent nodes using ROADMs (OB reduces the amount of QKD modules per path but incurs significant attenuation \cite{zhang2023routing}). OB and TR enable efficient key routing over paths connecting non-adjacent nodes, while QKPs add a further degree of freedom to key distribution, facilitating key retrieval between node pairs that can be represented via auxiliary paths. Note that these technologies represent logical capabilities rather than specific physical components. 
Fig.~\ref{fig:keyallocation} shows four key requests. Request 1, between nodes (1,3), is served using cached QKD keys and marked with a purple line. Request 2, between nodes (2,4), is served via trusted relay and shown in orange. Requests 3 and 4, also between (1,3) and (2,4), are served via bypass and represented by green and blue lines, respectively. 
Fig. \ref{fig:keyallocation} also illustrates how a single attack on one physical link can disrupt multiple requests, even those that traverse different links. Specifically, a high-power jamming attack on link (1,2) disrupts Request 3, but, since Request 3 uses OB, the attack causes the high-power signal to be propagated also to link (2,3), affecting Request 2. Since Request 1 uses pre-stored keys from the QKP, it remains unaffected.

To mitigate the impact of these jamming attacks, we formulate in this study the \textit{Routing and Wavelength Assignment with Minimal Attack Radius (RWA-MAR)} problem, which seeks to minimize the worst-case impact of a single physical-layer jamming attack. We propose a metric called maxNAR (Maximum Number of Affected Requests), i.e., an extension of the NAR (Number of Affected Requests) tailored to the unique characteristics of QKD systems. maxNAR captures the worst-case service disruption from a single compromised link, factoring in QKD-specific features such as wavelength sharing and specialized transmission mechanisms.

Our proposed approaches address the temporal dynamics of QKD networks by dividing time into discrete slots, during which key availability fluctuates based on the current QKP storage. Unlike conventional optical networks that utilize the Lightpath Attack Radius (LAR) metric \cite{furdek2011physical}, we adopt the NAR as a more suitable indicator for QKD scenarios.  

\vspace{-4mm}
\vspace{-5mm}

\subsection{Related work}
\label{sec:related-work}

QKD networks have been successfully demonstrated in testbeds in various countries, including Switzerland and Italy \cite{cao2022evolution}, demonstrating their potential for secure communication. The feasibility of co-existence between QKD and conventional communication signals in optical fiber networks has also been validated \cite{gatto2021bb84}. A QKD network architecture is commonly organized into three hierarchical layers: the Quantum Point-to-Point (or Link) layer, the Quantum Network Layer, and the Quantum Transport Layer \cite{dianati2008architecture}. Within this framework, keys can be generated and distributed through QKD nodes and modules, utilizing advanced technologies such as TR and OB \cite{cao2018time}. TR is a foundational component in many existing QKD testbeds \cite{cao2022evolution}, enabling long-distance key distribution, including a demonstration extending up to 4,600 kilometers \cite{chen2021integrated}. OB has been explored both experimentally and analytically \cite{brunner2023demonstration}. The authors of \cite{dong2020auxiliary} proposed a quantum-node representation in graphs that facilitates OB using auxiliary graphs to capture quantum-node logical adjacencies enabled by OB. The authors of \cite{zhang2023routing} examined OB and TR integration through a network-wide optimization lens. In our prior work \cite{li2024drl}, we employed a combination of OB and TR with QKP to support progressive recovery from large-scale network disruptions.

With the growing deployment of QKD networks, an important next research question is how to ensure QKD networks resiliency against physical-layer attacks.
Physical-layer attacks can be launched against QKD networks, potentially disrupting key generation~\cite{smith2021out, alomari2024securing, mani2024security}. While resilience to such attacks has been studied in classical optical networks~\cite{furdek2011physical}, solutions specifically tailored to the unique properties of QKD networks remain unexplored.  

In contrast to classical optical networks, the presence of technologies introduced for QKD networks, such as QKP caching, OB, and TR, significantly reshapes how disruption propagates. In this paper, we propose a new approach for solving the attack-aware routing problem in QKD networks.

\subsection{Contribution}

The main technical contributions of this work can be summarized as follows:
\begin{itemize}
    \item We formulate a new problem, RWA-MAR, as an ILP, and define a new maXNAR metric to quantify the worst-case impact of a single physical-layer attack. 
    \item We develop a scalable heuristic algorithm and a baseline method, to solve large-scale maxNAR instances efficiently where ILP is computationally infeasible.
    \item We provide comprehensive numerical evaluations to assess the impact of OB/TR architectures on security and resource utilization. 
\end{itemize}

The rest of the paper is organized as follows. Section~\ref{sec:model} formally defines the problem definition, its ILP formulation, and heuristic solution approach. Section~\ref{sec:result} presents and discusses simulation results under various architectural configurations. Finally, Section~\ref{sec:conclusion} concludes the paper.


\section{RWA-MAR problem}
\label{sec:model}
\subsection{System model}

We model the QKD network as a weighted directional graph $G_p = (N_p, E_p)$, where $N_p$ and $E_p$ are the sets of nodes and links, respectively. maxNAR as the maximum number of requests any one request is link-sharing with, where link-sharing is defined as a property indicating whether two requests traverse at least one common physical link in the same direction. We assume time is divided in timeslots. We then construct a fully-connected auxiliary graph $G_p = (N_p, E_a)$ where each link denotes the opportunity for key distribution between adjacent and non-adjacent nodes. Key distribution between adjacent nodes can use a physical quantum channel or a logical auxiliary link enabled by QKP. A quantum channel is where qubits are transmitted at different wavelengths.  Key distribution between non-adjacent nodes can use a quantum channel with OB/TR or an auxiliary link enabled by QKP key caching (i.e., stored keys in advance for future use). Each node is equipped with a limited number of QKD modules, including both transmitters and receivers. An OB connection requires 2 modules in total, whereas a TR path consumes two modules (a QKD receiver and a QKD transmitter) at each intermediate node to enable key forwarding. Our evaluations are grounded in realistic QKD key rate models \cite{zhang2023routing}, and incorporate both OB and TR mechanisms within the system architecture.


\begin{figure}[htbp]
\centering
\subfigure[NAR for TR, with maxNAR=3 ]{
\begin{minipage}[t]{0.4\textwidth}
\centering
\includegraphics[width=6.5cm, height=1.5cm]{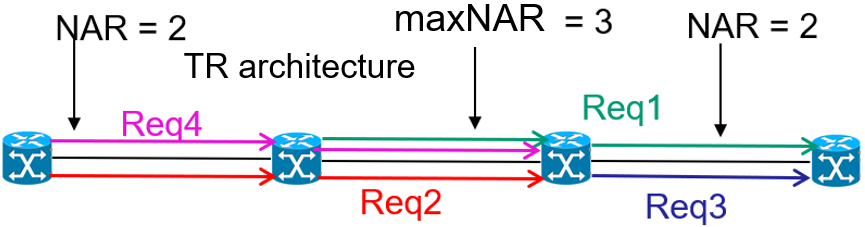}
\end{minipage}
}
\subfigure[NAR for OB, with maxNAR=4]{
\begin{minipage}[t]{0.4\textwidth}
\centering
\includegraphics[width=6.5cm, height=1.8cm]{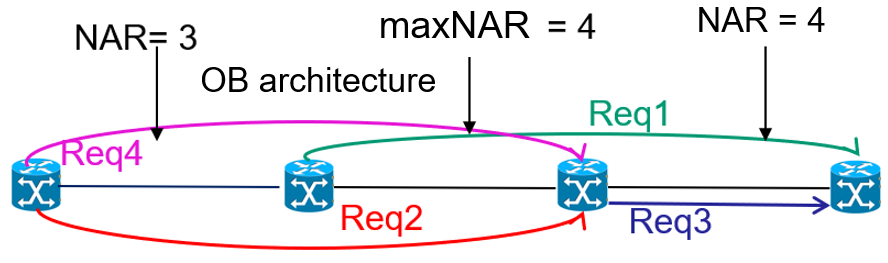}
\end{minipage}
}

\setlength{\abovecaptionskip}{-1mm}
\caption{Example of NAR calculation}
\vspace{-7mm}
\label{fig:NAR}
\end{figure}

Figure~\ref{fig:NAR} illustrates how NAR is computed under OB and TR architectures. Arrows indicate the position of an attack and the resulting number of affected requests. The maximum value among them defines the maxNAR.
In Fig. \ref{fig:NAR}(a), all the requests are served using TRs. Due to the characteristics of TR, in this case, the NAR only calculates how many requests traverse the same physical link.  TR consistently results in a low NAR (in this case, maxNAR is equal to 3), but requires a higher amount of QKD modules. 
In Fig. \ref{fig:NAR}(b), requests 1, 2, and 4 use OB, while request 3 is served on a quantum channel between two adjacent nodes. Requests using OB are vulnerable to attack propagation because OB allows signals to pass through multiple nodes without regeneration, enabling interference to spread across downstream links. If there is an attack on the middle link, it affects requests 1, 2, and 4, with OB propagating the impact to request 3, keeping the NAR at 4. Note that the quantum channels are directional; thus, an attack in the reverse direction will not affect the served requests. Since keys distributed via QKP are always unaffected by physical-layer attacks (as shown in Fig.~\ref{fig:keyallocation}), we omit further examples involving QKP.

\subsection{Achievable Key Rate and QKP Capacity}
\label{sec:QKP}

The achievable key rate in our model is derived from Ref.~\cite{zhang2023routing}. With this model, we can calculate the maximum achievable key rate for different reaches reported in Table~\ref{tab1}.   
These key rates decrease by 11\% for each crossed node when using optical bypass. The QKP capacity is estimated based on AES256 encryption in Cipher Block Chaining mode. Since each key can securely encrypt up to $2^{48}$ AES blocks (=36000 Tb), a 100-channel fiber (1 Tb/s per channel) requires a 256-bit key rotation every 360 seconds. Thus, the minimum QKP capacity per one-hour stage is 2560 bits \cite{li2024drl}.

\vspace{-6mm}
\begin{table}[htbp]
\setlength{\abovecaptionskip}{-1mm}
\caption{Key rate for different reaches}
\begin{center}
\begin{tabular}{|c|c|c|c|c|c|}
\hline
Reaches&10km&20km&30km&40km&50km \\
\hline
Key rate&23 kb/s&13 kb/s&7 kb/s&3.5 kb/s &1.9 kb/s\\
\hline
\end{tabular}
\label{tab1}
\end{center}
\end{table}
\vspace{-8mm}

\subsection{Problem Statement}

The RWA-MAR problem can be stated as follows: \textbf{Given} a QKD network topology, number of QKD modules and quantum channels, key requests, achievable key rates for different reaches, timeslots,  \textbf{decide} the routing, wavelength, and key-rate assignment for key requests, \textbf{constrained} by the maximum achievable key rates on given paths, the maximum number of quantum channels and quantum modules, with the \textbf{objective} to minimize the sum of the maxNAR across all timeslot.
We solve the RWA-MAR problem under three network architectures depending on the availability of optical bypass and trusted relay: 1) OB-TR, 2) OB, 3) TR.

\subsection{Integer Linear Programming (ILP) model}
The sets, parameters, and variables for the ILP model are listed in TABLE II.\\

\vspace{-3mm}
\begin{table}[H]
\begin{footnotesize}
\caption{Sets, Parameters, and Variables Description for ILP Model.}
\begin{tabularx}{\linewidth}{|p{1cm}|X|}
\hline
\textbf{Sets}& \textbf{Description} \\
\hline
$G_p$ & Physical network topology\\
$N_p$ & Set of physical nodes in network\\
$E_p$ & Set of physical links in network\\
$E_a$ & Set of links in the fully-connected graph\\
$P$ & Set of node pairs of requests in the network\\
$W$ & Set of QKD channels\\
$T$ & Set of timeslots\\
$\Phi_e$ & Set of physical routes that use the same end nodes as auxiliary link $e \in E_a$\\
$D$ & Set of requests\\
$S^+(i)$ & Set of outgoing links from node $i$\\
$S^-(i)$ & Set of incoming links for node $i$\\
\hline

\end{tabularx}

\begin{tabularx}{\linewidth}{|p{1cm}|X|}
\hline
\textbf{Para}& \textbf{Description} \\
\hline
$O_n$ & Integer, number of QKD modules on node $n$\\

$h_{\phi,e}$ & Binary, equals to 1 if link $e \in E_p$ in the route $\phi$\\

$k_d$ & Integer, required key rate of request $d \in D$\\

$l_\phi$ & Integer, key rate that can be supplied by route $\phi$\\
\hline
\end{tabularx}

\begin{tabularx}{\linewidth}{|p{1cm}|X|}
\hline
\textbf{Var}& \textbf{Description} \\
\hline
$f_{e,w}^{p,t}$ & Binary, equals to 1 if quantum channel $w$ on link $e\in E_a$ is allocated for path between node pair $p$ at timeslot $t$\\

$x_{e,w}^{p,t,\phi}$ & Binary, equals to 1 if route $ \phi$ is selected for connection between the end nodes of link $e \in E_a$ in QKD path between node pair $p$ at channel $w$ at timeslot $t$\\

$xx^{p,t}_{e'}$ & Binary, at timeslot t, the routing of node pair p used physical link $e'$ in $E_a$\\

$p^{p,t}_{e,w}$  & Binary, equals to 1 if link $e \in E_a$ used quantum channel in between node pair $p \in P$ on channel $w \in W$ at timeslot $t \in T $\\

$q^{p,t}_{e,w}$ & Binary, equals to 1 if path $p$ contains auxiliary link $e \in E_a$ based on QKP on channel $w$ at timeslot $t$\\

$u_{p,w}^t$ & Integer, key rate generated for path $p$ on channel $w$ at timeslot $t$\\

$z_{p,w}^t$ & Binary, equals t if QKD path between node pair $p \in P$ uses quantum channel $w \in W$ at timeslot $t$\\

$B^{t}_{\phi}$ & Binary, at timeslot \textit{t}, the routing $\phi$ has been used\\

$C^{p,t}_{\phi}$ & Binary, at timeslot \textit{t}, any attack on routing $\phi$ will affect the routing of node pair $p$\\

$g^t_p$ & Integer, stored keys in QKP for path between node pair $p$ at timeslot $t$\\

$y_d^t$& Binary equals to 1 if request $d$ is served at timeslot $t$\\

$\gamma_{e,w}^{p,t}$ & Integer, key rate provided from QKP for link $e$ in QKD path between node pair $p$ in channel $w$\\

$maxNAR_t$ & Integer, the maxNAR of timeslot t\\
\hline
\end{tabularx}
\end{footnotesize}
\end{table}
\vspace{-3mm}

\textbf{Objective function:} to minimize the maxNAR.

\vspace{-2mm}
\begin{equation}
min\sum_{t\ \in\ T}maxNAR_t
\label{min-max}
\end{equation}\\
\vspace{-8mm}

\subsubsection{Flow, link, and modules constraints}
Eqs.~(\ref{flow-constraints}), (\ref{f-q-p}) show the flow constraint for the QKD path, and it can be either a quantum channel (OB/TR) or an auxiliary link (QKP enabled link). Eq.~(\ref{module-constraint}) ensures the number of used modules is smaller than or equal to the number of available nodules in a node.
For each link in a fully connected graph $E_a$, it may consist of several physical links $e \in E_p$ in physical topology.
Eq.~(\ref{x-q}) determines the physical route $\phi \in \Phi$ for auxiliary links $e \in E_a$.
Eqs.~(\ref{x-1}) and (\ref{x-h-1}) ensure that multiple QKD paths cannot use the same channel and the same route.

\vspace{-2mm}
\begin{equation}
\setlength\abovedisplayskip{-3pt}
    \begin{split}
    \sum_{e \in S^+(i)} f_{e,w}^{p,t} -\sum_{e \in S^-(i)}f_{e,w}^{p,t} 
    = 
    \begin{cases}
    z_{p,w}^t \  if\ i=a(p)\\  
    -z_{p,w}^t \ if\ i=b(p)\\
    0 \ others\\ 
    \end{cases}
    \\
    \quad\forall p \in P,i \in N_p, t \in T, w \in W 
\label{flow-constraints}
\end{split}
\end{equation}
\vspace{-5mm}

\vspace{-1ex}
\begin{equation}
f_{e,w}^{p,t} = q_{e,w}^{p,t} \lor p_{e,w}^{p,t}\ \ \forall e \in E_a, p \in P, t \in T, w \in W 
\label{f-q-p}
\end{equation}
\vspace{-7mm}

\vspace{-1ex}
\begin{equation}
\begin{split}
\sum_{p \in P,e \in S^+(n),w \in W}p_{e,w}^{p,t} +
\sum_{p \in P, e \in S^-(n), w \in W}p_{e,w}^{p,t}\\
\quad\leq O_n
\quad\forall n \in N_p, t \in T 
\label{module-constraint}
\end{split}
\end{equation}
\vspace{-3mm}

\vspace{-1ex}
\begin{equation}
\sum_{\phi \in \Phi_e}x_{e,w}^{p,t,\phi} = q_{e,w}^{p,t} \ \ \ \ \forall p \in P, e \in E_a, w \in W, t \in T
\label{x-q}
\end{equation}

\vspace{-1ex}
\begin{footnotesize}
\begin{equation}
\sum_{p \in P}x_{e,w}^{p,t,\phi} \leq 1 \ \ \ \ \forall e \in E_a, t \in T, w \in W,\phi \in \Phi_e
\label{x-1}
\end{equation}
\end{footnotesize}
\vspace{-3mm}

\vspace{-1ex}
\begin{equation}
\sum_{p \in P, e' \in E_a, \phi \in \Phi_e}x_{e,w}^{p,t,\phi} * h_{\phi,e'} \leq 1 
 \forall e \in E_p, w \in W, t \in T
\label{x-h-1}
\end{equation}
\vspace{-3mm}
\subsubsection{Key rate constraints} Eq.~(\ref{u-x-l-y-m-1-f}) ensures the key rate of QKP path $p$ is less than the sum of the key rate provided by a quantum channel and from QKP in each edge of the path.
Eq.~(\ref{u-m-z}) ensures that keys are distributed only when the corresponding path $p$ is active and available for use.

\vspace{-1ex}
\begin{equation} 
\begin{split}
u_{p,w}^t \leq \sum_{\phi \in \Phi_e}(x_{e,w}^{p,t,\phi}* l[\phi]) + \gamma_{e,w}^{p,t} + \\
M * (1 - f_{e,w}^{p,t})
 \ \forall e \in E_a, p \in P, w \in W, t \in T 
\label{u-x-l-y-m-1-f}
\end{split}
\end{equation}
\vspace{-3mm}

\vspace{-1ex}

\begin{equation}
u_{p,w}^t \leq M * z_{p,w}^t \ \ \ \ \forall p \in P, t \in T, w \in W 
\label{u-m-z}
\end{equation}
\vspace{-3mm}

\subsubsection{QKP storage constraints}  Eq.~(\ref{g-0}) ensures that the stored keys in QKP are not less than 0. Eq.~(\ref{g-g-u-y-y-k-y}) expresses that the amount of keys stored in the QKP at stage $t$ equals the remaining keys from the previous stage $t-1$, plus the newly generated keys supplied by the quantum channel, minus the keys consumed by all requests routed through path $p$. Eq.~(\ref{y-m-p}) ensures that the variable $\gamma_{e,w}^{p,t}$ equals to 1 only when QKP is being used for path $p \in P$.

\vspace{-3mm}
\begin{equation}
g^t_p \geq 0 \ \ \ \ \forall p \in E_a, t \in T
\label{g-0}
\end{equation}

\vspace{-1ex}
\begin{equation}
\begin{split}
 g^t_p \leq g^{t-1}_p + \sum_{w \in W}u_{p,w}^t - \sum_{p' \in E_a}\sum_{w \in W}(\gamma_{p,w}^{p',t} + \gamma_{\bar{p},w}^{p',t})\\
 - k_p*y^t_p
 \ \forall p \in P, t \in T
 \label{g-g-u-y-y-k-y}
 \end{split}
 \end{equation}
 
\vspace{-1ex}
 \begin{equation}
\gamma_{e,w}^{p,t} \leq M * q_{e,w}^{p,t} \ \ \ \ \forall p \in P, e \in E_a, t \in T, w \in W
\label{y-m-p}
 \end{equation}
\vspace{-3mm}

\subsubsection{NAR constraints} Eq.~(\ref{xx-x}) ensures that the variable $xx^{p,t}_{e'}$ equals to 1 when path $p$ uses physical link $e'$. 
Eq. (\ref{b-x}) defines if route $\phi$ is used on timeslot $t$.
Eq.(\ref{c-x}) ensures route $\phi$ has been used for request $d$ at timeslot $t$.
Eq.(\ref{maxlar}) calculates NAR at each timeslot $t$. 

\vspace{-3mm}
\begin{equation}
\begin{split}
xx^{p,t}_{e'} \geq x^{p,t,\phi}_{e,w} \cdot h_{\phi, e'}\forall t\in T,  w\in W, p\in P,\\
e\in E_a, e'\in E_p, \phi \in \Phi_e
\label{xx-x}
\end{split}
\end{equation}

\vspace{-3mm}
\begin{equation}
B^{t}_{\phi} \geq x^{p,t,\phi}_{e,w} \forall\ t\ \in\ T, w\ \in\ W, p\ \in\ P, e\ \in\ E_a, \phi\ \in\ \Phi_e
\label{b-x}
 \end{equation}
 \vspace{-3mm}
 
\begin{equation}
\begin{split}
C^{d,t}_{\phi} \geq \left(xx^{d,t}_{e'} \cdot h_{\phi, e'}\right)\ \land \ B^{t}_{\phi} \\
\forall\ t\in T,  d\in D, e'\in E_p, e\in E_a,\phi\in\Phi_{e}
\label{c-x}
\end{split}
\end{equation}

\vspace{-1mm}
\begin{equation}
maxNAR_t \geq \sum_{d\ \in\ D} C^{d,t}_{\phi}\ \forall\ t\ \in \ T,    e\ \in\ E_a,   \phi\ \in\ \Phi_{e}
\label{maxlar}
\end{equation}
\vspace{-3mm}



\subsection{Min-maxNAR Algorithm}

\begin{algorithm}[h]
\footnotesize
    \caption{Min-maxNAR Algorithm}
    \textbf{Input:}{$N_p$, $E_p$, $G_p$, $T$, $D$, $k_d$, $J^p_d$, $\alpha$, $W$, $O_n$}\\
    \textbf{Output:}{maxNAR, average NAR}\\
    \begin{algorithmic}[1]
                \For {$t \in T$. At timeslot t=0, input $J^p_d$ = 0 \do}
                    \For {each unserved request $J^p_d < k_d$, $d$ = 1 to $\lvert D \rvert$ \do}
                        \State {Get shortest path $P_d$ for all the requests. For OBTR architecture, set a tunable parameter $\alpha\%$ to select the initial shortest path.}
                        \If {Path $P_d$ for request $d$ routing from $d_s$ to $d_d$ exists.}
                                \State {Distribute keys $g_d^p$ from path $P_d$, and stored keys in QKP }
                                \State {Update quantum channel and modules utilization.} 
                                \State {$J^p_d = J^p_d + g_d^p$} 
                        \EndIf
                    \EndFor

                    \While {QKPs is not full and network resources are not exhausted\do}
                            \State {Find the routing path and store more keys for the future}
                            \State {Update quantum channel and QKD modules utilization}
                    \EndWhile
                    \For {iteration $\leq \theta$ \do}
                        \State {Randomly select a lightpath to reroute}
                        \State {Find the K-shortest paths as the neighborhood of the lightpath}
                        \State {Find the best neighbor which is not in the Tabu list}
                        \State {Update quantum channel and modules utilization}
                        \State {Update the maxNAR, and average NAR}
                    \EndFor
                \EndFor
                \State{return maxNAR, and average NAR}
    \end{algorithmic}
    \label{qmnar}
\end{algorithm}

To solve the RWA-MAR problem, we developed a scalable heuristic algorithm. 
The details of the Min-maxNAR algorithm are reported in Alg.\ref{qmnar}. Min-maxNAR algorithm constructs a topology $G_t$ for each stage, which contains nodes and links. $J^d_p$ is an array whose elements contain the achievable key rate from path $p$ for each request $d$. We develop the Min-maxNAR Algorithm based on the Tabu Search Algorithm (TSA) \cite{glover1990tabu}, as  TSA has proven powerful in solving min-max problems like ours\cite{youssef2001evolutionary}.

At lines 1-2, the topology is initialized, and the system checks the current time-slot while identifying all unserved requests. From lines 3-9, the algorithm first finds the shortest path as the initial routing for serving all the requests from $d = 1$ to $|D|$. For OBTR architecture, initial paths are selected using TR or OB based on a tunable priority parameter $\alpha$. A lower $\alpha$ favors OB for reduced module cost at the expense of a slightly higher maxNAR, while a higher $\alpha$ prioritizes TR to minimize maxNAR, accepting higher resource usage. OBTR0 refers to the OBTR architecture with $\alpha = 0$, meaning the algorithm initially prioritizes OB paths during path selection and stores them in the Tabu list as the starting solution. Even under OBTR0, however, the system can still have the possibility to select TR during Tabu-search iterations. Note that $\alpha$ only influences the initial solution; afterward, the heuristic uses Tabu-Search to iteratively optimize routing, often incorporating more TR paths to improve maxNAR. 
Then the algorithm stores the distributed keys in the QKP while updating the network resource usage, including quantum channels and modules. Moving to lines 10-13, the algorithm examines whether redundant resources are available to serve future requests, allowing for storage of keys in advance using the QKP, following the same logic applied in lines 3 to 5. In lines 14–21, the heuristic optimizes routing by exploring neighboring solutions of the current path, selecting the one with the lowest maxNAR, and updating both the TABU list and the resource utilization (quantum channels and module) accordingly.

\section{Illustrative Numerical Results}
\label{sec:result}
\vspace{-3mm}
\begin{figure}[H]

  \centering
  \includegraphics[width=0.4\textwidth]{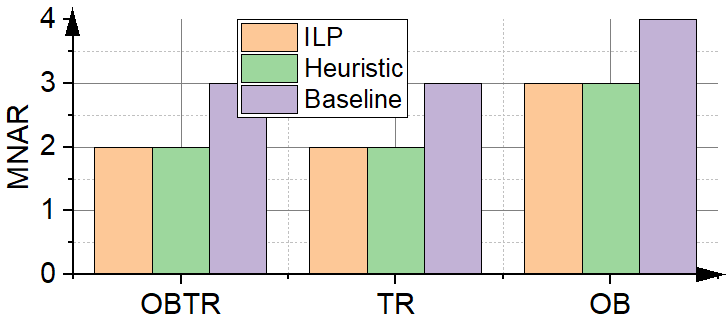}
  \setlength{\abovecaptionskip}{-1mm}
  \caption{Results for PoliQi topology.}
  \label{fig:ilp}
\end{figure}
\vspace{-4mm}

We evaluate the performance of the Min-maxNAR algorithm under three different architectures (\textit{OB-TR}, \textit{OB}, and \textit{TR}).
We first benchmark the performance of the in-maxNAR algorithm compared to the ILP in a five-node ring topology (as in the PoliQi QKD testbed currently being deployed in Milan \cite{zhang2023routing}). Each node has ten modules. We consider 7 requests, and each request requires 10kb/s during each timeslot.  We then evaluate the Min-maxNAR algorithm also in the NSF topology \cite{dong2012energy}, which has 14 nodes (70 modules) and 21 links. We scale down the link distance as [5, 15]km to be suitable for QKD reaches. Due to the lack of scalability, ILP can not be solved for the NSF topology. For the NSF topology, we consider a demand matrix in which requests are generated for 80\% of all node pairs. 80\% of the requests have key rate distributed in [5-10] kb/s, and 20\% of requests have key rate distributed in [15-25] kb/s. We evaluate our heuristic’s performance based on three key metrics: maxNAR, average NAR (avgNAR), and average modules cost for each node, comparing it against a baseline that utilizes depth-first search for shortest path routing between node pairs in the network graph. Note that we also use a tunable parameter $\alpha$ to adjust the initial TABU list in our Min-maxNAR algorithm.

We first discuss results on the PoliQi topology. As shown in Fig.~\ref{fig:ilp}, the proposed heuristic achieves the same (optimal) maxNAR of the ILP, with OB-TR and TR architectures reaching a maxNAR of 2 (note that TR consumes more QKD modules than OB-TR, and here $\alpha$ = 0). As expected, OB yields the highest maxNAR. Notably, the ILP requires over ten hours to converge, while the heuristic achieves the same results in about five seconds.

\begin{figure*}[htbp]
\centering
\subfigure[maxNAR ]{
\begin{minipage}[t]{0.31\textwidth}
\centering
\includegraphics[width=5.3cm,height=4.1cm]{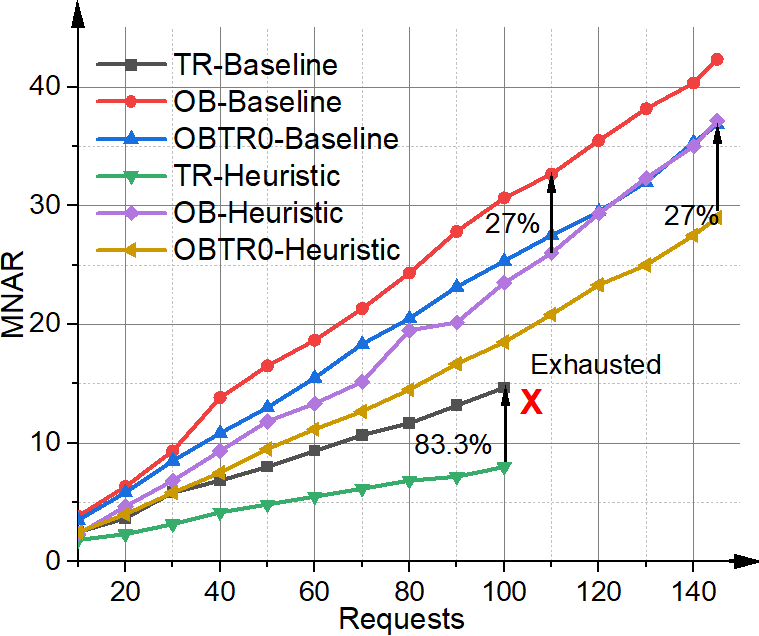}
\end{minipage}
}
\subfigure[avgNAR]{
\begin{minipage}[t]{0.31\textwidth}
\centering
\includegraphics[width=5.3cm,height=4.1cm]{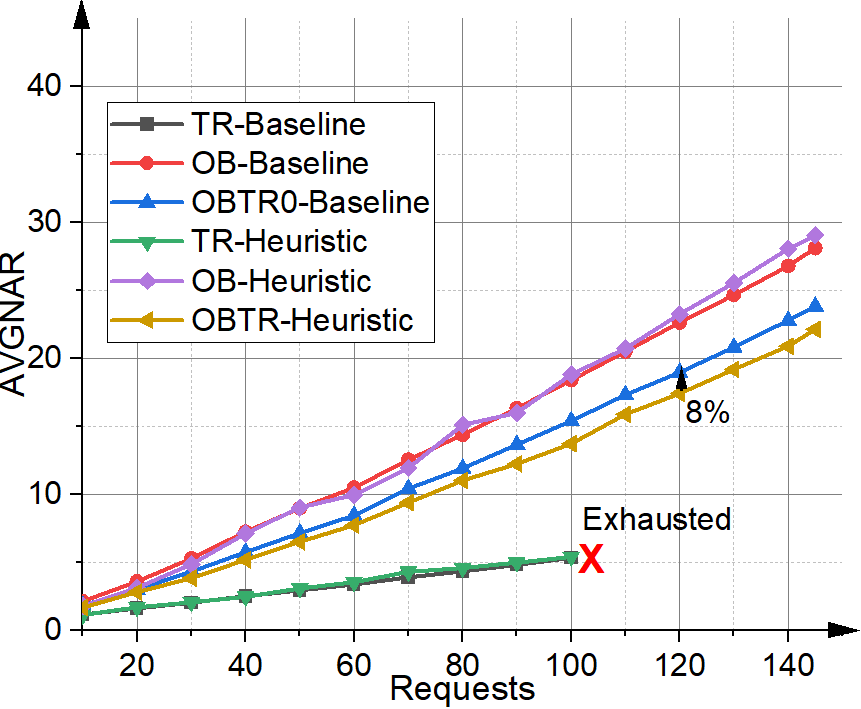}
\end{minipage}
}
\subfigure[Average number of cost modules for each node]{
\begin{minipage}[t]{0.31\textwidth}
\centering
\includegraphics[width=5.3cm,height=4.1cm]{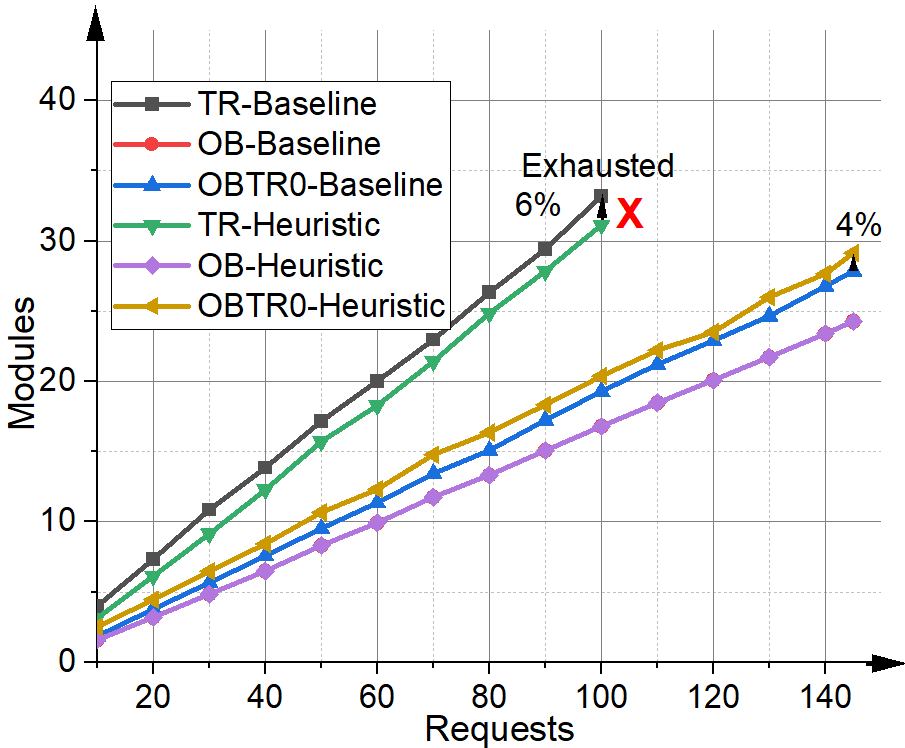}
\end{minipage}
}
\setlength{\abovecaptionskip}{-1mm}
\caption{Result in NSF topology: compared with baseline with $\alpha = 0$}
\vspace{-4mm}
\label{fig:14-1}
\end{figure*}

Next, we consider the NSF topology and we analyze the performance of the Min-maxNAR algorithm compared to the baseline across the three architectures. In Fig.~\ref{fig:14-1}(a), for OB and OBTR0 configurations (meaning $\alpha$ = 0), the heuristic reduces maxNAR by approximately 27\% compared to the baseline. In contrast, TR exhausts resources after 100 requests due to its higher module consumption. As expected, OB exhibits the highest maxNAR, while TR achieves the lowest.
Figure \ref{fig:14-1}(b) presents the avgNAR, which represents the average NAR across all physical links and requests in the network. Compared to the maxNAR results, the gap between the heuristic and baseline approaches is narrower. For both OB and TR architectures, the heuristic yields a higher avgNAR. This is because the heuristic reduces the maxNAR by distributing the requests more evenly across the network, which in turn slightly raises
the average number of affected requests. However, in the OBTR architecture, the flexibility of the heuristic allows for a more optimized allocation, resulting in both lower maxNAR and avgNAR compared to the baseline. Specifically, the heuristic achieves up to 8\% reduction in avgNAR.
Among the architectures, TR consistently achieves the lowest avgNAR, while OB results in the highest, as expected.
Finally, Fig.~\ref{fig:14-1}(c) illustrates module utilization. TR consumes the most modules, and OB the least (only costs two per lightpath). For OB, both heuristic and baseline use the same number of modules, as expected. In TR, the baseline uses about 6\% more modules than the heuristic, while in OBTR, the heuristic consumes 4\% more modules to achieve better maxNAR and avgNAR, reflecting a deliberate trade-off for improved routing flexibility. 

\begin{figure*}[htbp]
\centering
\subfigure[maxNAR ]{
\begin{minipage}[t]{0.31\textwidth}
\centering
\includegraphics[width=5.3cm,height=4.1cm]{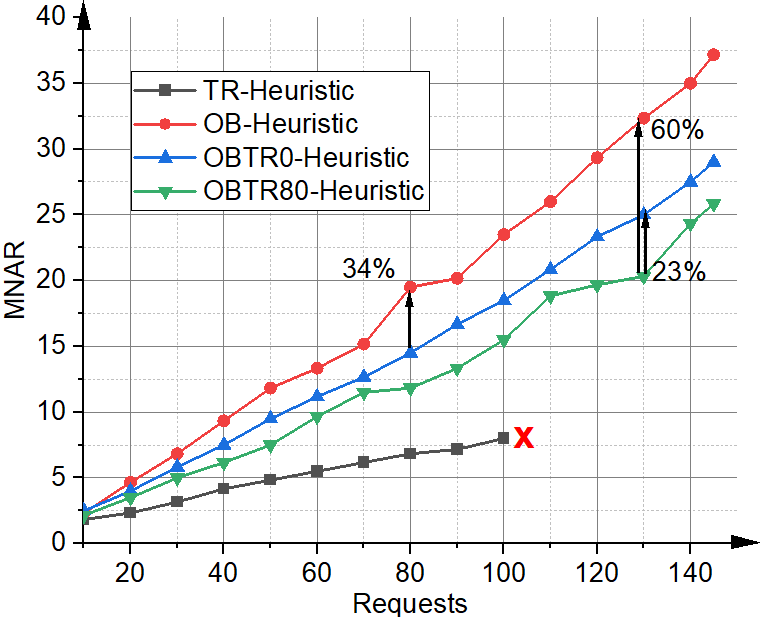}
\end{minipage}
}
\subfigure[avgNAR]{
\begin{minipage}[t]{0.31\textwidth}
\centering
\includegraphics[width=5.3cm,height=4.1cm]{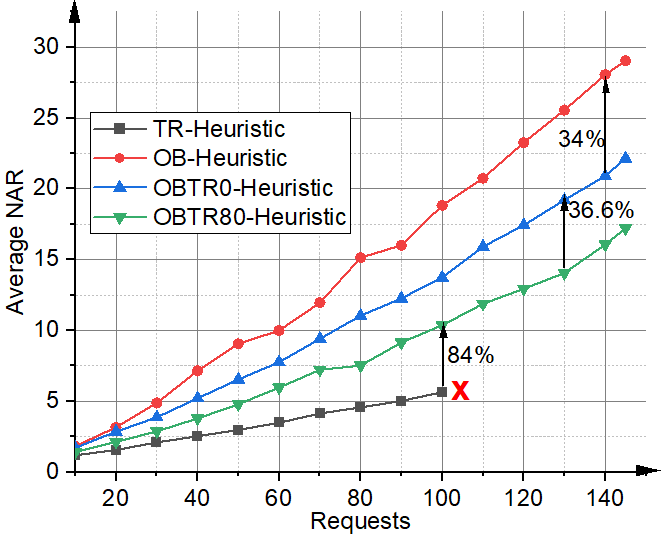}
\end{minipage}
}
\subfigure[Average number of cost modules for each node]{
\begin{minipage}[t]{0.31\textwidth}
\centering
\includegraphics[width=5.3cm,height=4.1cm]{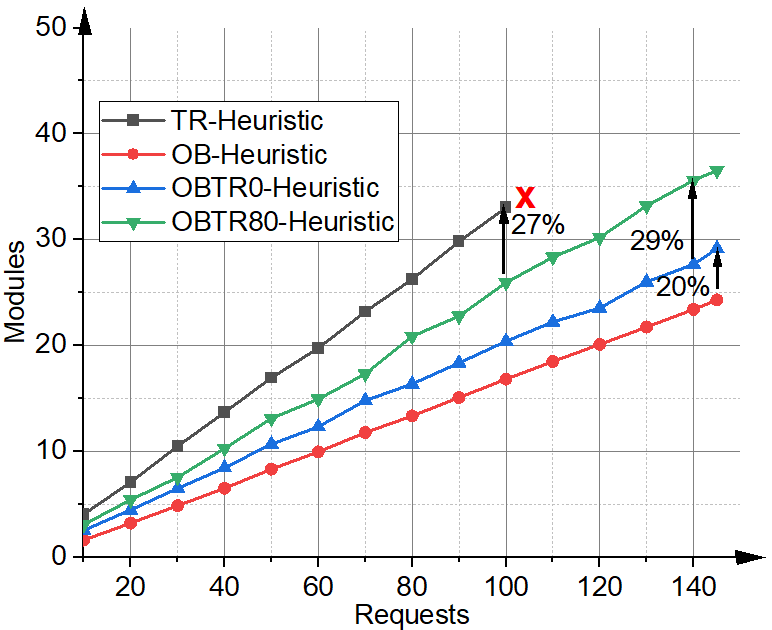}
\end{minipage}
}
\setlength{\abovecaptionskip}{-1mm}
\caption{Result in NSF topology with $\alpha = 0$ $\And$  $\alpha = 80$}
\vspace{-6mm}
\label{fig:14-2}
\end{figure*}

Moreover, we analyze the impact of the priority parameter $\alpha$ on maxNAR, avgNAR, and module utilization, as shown in Fig.~\ref{fig:14-2}. OBTR80 corresponds to $\alpha=80$, meaning an 80\% preference for initially selecting TR paths, while OBTR0 fully prioritizes OB paths. As shown in Fig. \ ref {fig:14-2}(a), OBTR80 achieves a 23\% maxNAR reduction compared to OBTR0 and a 60\% reduction relative to OB. Even OBTR0 provides a 34\% improvement over OB. In terms of avgNAR (Fig.\ref{fig:14-2}(b)), OBTR80 outperforms OBTR0 by 36.6\%, while OBTR0 achieves a 34\% gain over OB. TR consistently yields the lowest avgNAR, exhibiting an 84\% gap compared to OBTR80.
Module utilization results are shown in Fig.~\ref{fig:14-2}(c). TR exhausts resources due to its heavy module demand, while OBTR80 consumes 27\% fewer modules than TR but 29\% more than OBTR0. OBTR0 increases module utilization by 20\% compared to OB, while achieving maxNAR and avgNAR gains. These results show that $\alpha$ is a critical tuning parameter: lower $\alpha$ is preferred with limited modules, while higher $\alpha$ improves security and costs more resources.

\begin{figure}[htbp]
  \centering
  \includegraphics[width=0.45\textwidth]{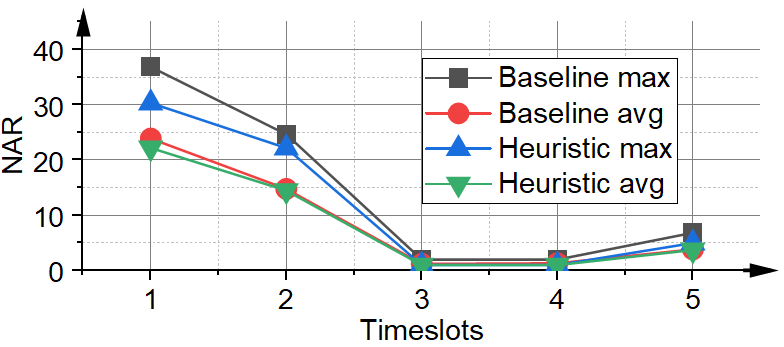}
    \setlength{\abovecaptionskip}{-2mm}
  \caption{Result for 5 timeslot with 145 requests}
  \label{fig:5}
\end{figure}

Finally, we evaluate the impact of QKPs across multiple timeslots using the OBTR0 configuration in the NSF topology with 145 requests. As shown in Fig.~\ref{fig:5}, both baseline and heuristic start with high maxNAR, but the heuristic reduces it from 37 to 32, while avgNAR shows smaller differences. The usage of key caching leads to a sharp drop in NAR during the second timeslot. By the third and fourth timeslots, maxNAR stabilizes around 1 to 2 as QKP reserves meet demand. A slight rise occurs in the final slot as keys are nearly depleted, but the heuristic still outperforms, demonstrating efficient dynamic key allocation.

\vspace{-3mm}
\vspace{-5mm}

\section{Conclusion}
\label{sec:conclusion}

This paper addresses QKD network vulnerability to physical-layer attacks by formulating the RWA-MAR problem and developing both an ILP model and a scalable TABU-based heuristic to solve it. The ILP and heuristic are applicable in three different technological scenarios, namely, OB, TR, and OBTR. Simulation results demonstrate that our heuristic outperforms a baseline solution by about 27\% in terms of maxNAR and average NAR. To our knowledge, this is the first work to model and optimize maxNAR for improving QKD network resilience.


\section*{Acknowledgment}
\renewcommand{\baselinestretch}{0.5}
This work was supported in part by funding from the Innovation for Defence Excellence and Security (IDEaS) program from the Department of National Defence (DND). 
This work was partially supported by the PRIN project ZeTON funded by Italian Ministry of University and Research.

\renewcommand{\baselinestretch}{0.9}
\bibliographystyle{ieeetr}
{\footnotesize
\bibliography{NAR}}

\end{document}